\documentclass{elsart}
\journal{Physica D}
\usepackage{amsmath} \usepackage{bm} \usepackage{graphicx}
\usepackage{amsfonts} \usepackage{amssymb}
\begin{document}
\def\phi{\varphi} \def\epsilon{\varepsilon} \def\u{\bm}
\def\Bbb{\mathbb}

\begin{frontmatter}
\title{Determination of the threshold of the break-up of invariant
    tori in a class of three frequency Hamiltonian systems}
  
  \author{C.\ Chandre}
 \address{Center for Nonlinear Science, School of Physics, Georgia Institute of Technology,
 Atlanta, GA 30332-0430, USA}
 \author{J.\ Laskar}
\address{Astronomie et
    Syst\`emes Dynamiques, IMC, CNRS UMR8028, 77 Avenue 
    Denfert-Rochereau, F-74014 Paris, France}

 \author{G.\ Benfatto}
\address{Dipartimento di
    Matematica, Universit\`a di Roma ``Tor Vergata'', Via della
    Ricerca Scientifica, I-00133 Roma, Italy}

\author{H.R.\ Jauslin}
  
  \address{Laboratoire de Physique, CNRS, Universit\'e de
    Bourgogne, B.P. 47 870, F-21078 Dijon, France}

\begin{abstract}
  We consider a class of Hamiltonians with three
  degrees of freedom that can be mapped into quasi-periodically driven
  pendulums. The purpose of this paper is to determine the
  threshold of the break-up of invariant tori with a specific
  frequency vector. We apply two techniques~: the frequency map
  analysis and renormalization-group methods.  The renormalization
  transformation acting on a Hamiltonian is a canonical change of
  coordinates which is a combination of a partial elimination of the
  irrelevant modes of the Hamiltonian and a rescaling of phase space
  around the considered torus.  We give numerical evidence that the
  critical coupling at which the renormalization
  transformation starts to diverge is the same as the value given by
  the frequency map analysis for the break-up of invariant tori.
  Furthermore, we obtain by these methods numerical values of the threshold of
  the break-up of the last invariant torus.
\end{abstract}
\end{frontmatter}

\section{Introduction}\label{sec.1}

For Hamiltonian systems, the persistence of invariant tori influences
the global properties of the dynamics.  The study of the break-up of
invariant tori is thus an important issue to understand the onset of
chaos. For two degrees of freedom, there are several numerical methods to
determine the threshold of the break-up of invariant tori~: for
instance, Greene's criterion~\cite{gree79}, obstruction
method~\cite{olve87}, converse KAM~\cite{mack85,mack89}, frequency map
analysis~\cite{lask90,lask92,lask93,lask99}, or renormalization-group
methods~\cite{govi97,chan98a,chan98b}.\\
In this article, we propose to compute this threshold for a
one-parameter family of Hamiltonians with three degrees of freedom and
for a specific frequency vector, by two techniques~: by frequency map
analysis and by renormalization.  The frequency map analysis is valid
for any dimension, and has been applied to systems with a large
number of degrees of freedom~\cite{lask90}.  The set-up of
renormalization-group transformations is also possible for any
dimensions in the framework of Ref.~\cite{koch99}, but only systems
with two degrees of freedom have been investigated
numerically.\\
We describe the renormalization-group transformation and we implement
it numerically for the spiral mean torus.  The result is that the
values of the critical coupling given by the renormalization coincide
up to numerical precision with the thresholds of the break-up of the
spiral mean torus (of dimension 3) given
by frequency map analysis. The two methods we compare are completely
independent, both conceptually and in their practical
realizations. The frequency map analysis is based on the analysis of
trajectories, while the renormalization is based on a criterion of
convergence of a sequence of canonical transformations.\\
We conjecture, on the basis of this numerical result, that the
renormalization-group transformation converges up to the critical
surface (the set of Hamiltonians where the torus of the given
frequency is critical, i.e.\ at the threshold of its break-up), at
least in a region of the critical surface of the Hamiltonian space
where critical couplings are small
enough (in order that the elimination procedure is well-defined~\cite{koch99b}).\\

We consider a class of Hamiltonians with three degrees of freedom written in
terms of actions $\u{A}=(A_1,A_2,A_3)\in{\Bbb R}^3$ and angles
$\u{\phi}=(\phi_1,\phi_2,\phi_3)\in {\Bbb T}^3$ (the 3-dimensional
torus parametrized, e.g., by $[0,2\pi]^3$)
\begin{equation}
\label{eqn:ham}
H_{\varepsilon}(\u{A},\u{\phi})=H_0(\u{A})+\varepsilon V(\u{\phi}),
\end{equation}
where $\varepsilon$ denotes the coupling parameter.  In this article,
we consider the particular class of models for which the integrable
part $H_0$ is given by
\begin{equation}
\label{eqn:h0}
H_0({\u A})={\u \omega}_0\cdot{\u A}+\frac{1}{2}({\u \Omega}\cdot{\u A})^2,
\end{equation}
where $\u{\omega}_0$ is the frequency vector of the considered
invariant torus, and $\u{\Omega}$ is another constant vector
non-parallel to $\u{\omega}_0$.  We suppose that $\u{\omega}_0$ is
incommensurate, i.e.\ there is no nonzero integer vector $\u{\nu}$
such that
$\u{\omega}_0\cdot\u{\nu}=0$.\\
Since the quantity $\u{\omega}_0^{\perp}\cdot\u{\varphi}$ is conserved
(where $\u{\omega}_0^{\perp}$ denotes a vector orthogonal to
$\u{\Omega}$ and to $\u{\omega}_0$), one can show (even if
$\u{\omega}_0^{\perp}\cdot\u
{\varphi}$ is not a function on the three-dimensional torus) that this
model
(\ref{eqn:ham})-(\ref{eqn:h0}) is intermediate between two and three
degrees of freedom; in appropriate coordinates it can be interpreted
as one degree of freedom driven by a multi-periodic force with
incommensurate frequencies $\u{\omega}_0$.  In particular, invariant
tori in this
intermediate model act as barriers in phase space (limiting the
diffusion of trajectories) in a similar way as for two degrees of
freedom Hamiltonian systems. We analyze in this article the break-up
of invariant tori with spiral mean frequencies
for this particular type of models, by choosing a special form of the
perturbation
(see section \ref{sect:result}), such that the model is equivalent to a
pendulum driven
by two periodic forces with incommensurate frequencies.
The method is however applicable to any perturbation and
to the case of full three degrees of freedom~\cite{koch99,chan99b}.\\
We are interested in the stability of the torus with frequency vector
$\u{\omega}_0$.  For the unperturbed Hamiltonian $H_0$, this torus is
located at $\u{A}=\u{0}$.  Kolmogorov-Arnold-Moser (KAM) theorems were
proven for Hamiltonians (\ref{eqn:ham}) provided that $\u{\omega}_0$
satisfies a Diophantine condition~\cite{chan98c}.  This theorem shows
the existence of the torus with frequency vector ${\u \omega}_0$ for a
sufficiently small and smooth perturbation $\varepsilon V$.  
The invariant torus
is a small deformation of the unperturbed one. The existence of the
torus outside the perturbative regime is still an open question even
if efforts have been made to increase lower bounds for specific models
(for a two dimensional model, see Ref.~\cite{cell88,cell00}).
Conversely, for sufficiently large values of the coupling parameter,
it has been shown that the torus does no longer exist \cite{mack85,mack89}.
The aim of this paper is to determine $\varepsilon_c$ such that
$H_\varepsilon$ has a smooth invariant torus of the given frequency
for $|\varepsilon|<\varepsilon_c$, and does not have this
invariant torus for $|\varepsilon|>\varepsilon_c$.\\
The invariant torus we study (named the {\em spiral mean} torus) has
the frequency vector
$$
{\u \omega}_0=(\sigma^2,\sigma,1),
$$
where $\sigma$ is the spiral mean, i.e.\ the real root of
$\sigma^3=\sigma+1$ ($\sigma\approx 1.3247$).  From some of its
properties, $\sigma$ plays a similar role as the golden mean in the
two degrees of freedom case~\cite{kim86}.  The analogy comes from the
fact that one can generate rational approximants by iterating a {\em
  single} unimodular matrix $N$.  In what follows, we call {\em
  resonance} an element of the sequence $\{ {\u \nu}_k=N^{k-1}{\u
  \nu}_1, k\geq 1\}$ where ${\u \nu}_1=(1,0,0)$ and
$$
N=\left(
\begin{array}{ccc}
0 & 0 & 1 \\
    1 & 0 & 0\\
    0 & 1 & -1
            \end{array}\right).
          $$
          
          The word {\em resonance} refers to the fact that the small
          denominators ${\u \omega}_0\cdot {\u \nu}_k$ appearing in
          the perturbation series or in the KAM iteration, tend to
          zero geometrically as $k$ increases ($ {\u \omega}_0 \cdot
          {\u \nu}_k = \sigma^{3-k} \to 0 \mbox{ as } k\to \infty $).
          We notice that $\u{\omega}_0$ is an eigenvector of
          $\tilde{N}$, where $\tilde{N}$ denotes the transposed matrix
          of $N$.  One can prove~\cite{koch99} that $\u{\omega}_0$
          satisfies a Diophantine condition of the form~:
          $$
          |{\u \omega}_0\cdot {\u \nu}| >c |{\u \nu}|^{-2},
          $$
          where $|{\u \nu}|=(|\nu_1|^2+|\nu_2|^2+|\nu_3|^2)^{1/2}$,
          and $c\approx 0.6$.
\section{Renormalization-group transformation}\label{sec.2}

The renormalization transformations are defined for a {\em fixed}
frequency vector $\u{\omega}_0$, and contain a partial elimination of
the irrelevant modes (the non-resonant part) of the perturbation, and
a rescaling of phase space. The elimination of irrelevant modes is
performed by iterating a change of coordinates as in KAM theory. We
remark that other perturbative techniques can be used instead, leading
to similar results.  The rescaling of phase space combines a shift of
the resonances, a rescaling of time, and a rescaling of the actions.
The aim is to change the scale of the actions to a smaller one (and a
longer time scale).  This renormalization can be thought as a
microscope in phase space.  The non-resonant modes are the ones which
affect the motion at short time scales, and can be dealt with
averaging methods.  We define the non-resonant modes to be the ones
that satisfy the inequality
\begin{equation}
\label{eqn:cond}
\vert \u{\omega}_0\cdot\u{\nu}\vert > \frac{1}{\sqrt{2}}
\vert \u{\omega}_0\vert \vert \u{\nu}\vert.
\end{equation}
This set of modes, denoted $I^-$, is the interior of a cone around the
$\u{\omega}_0$-direction in the space of 3-dimensional vectors, with
angle $\pi/4$. We define the resonant modes as the Fourier modes which
do not satisfy the condition (\ref{eqn:cond}), i.e.\ this set, denoted
$I^+$, is the complement of $I^-$ in ${\Bbb Z}^3$.  Since $\u{\nu}_k$
does not satisfy Eq.~(\ref{eqn:cond}) for $k\geq 1$, $I^+$ contains
the resonances that produce small denominators in the perturbation
series or in the KAM theory.  The ``frequency cut-off'' (between
resonant and non-resonant modes) restricts the Fourier modes that can
be eliminated in one renormalization step, without running into small
denominator problems (the non-resonant modes). As it is common with
cut-offs, there is not a single ``natural'' choice.  More generally,
other choices in the splitting of $\{ e^{i\u{\nu}\cdot\u{\varphi}}\}$
into resonant and non-resonant modes should lead to the same results
provided, e.g., that the ratio $|{\u \nu}|/|{\u \omega}_0\cdot{\u
  \nu}|$ is bounded on $I^-$, and that the shift of the resonances
contracts non-zero vectors ${\u \nu}$ in $I^+$ and maps them into
$I^-$ after a finite number of iterations
of the transformation~\cite{chan98b,koch99}.\\

The transformation, acting on a Hamiltonian $H$ of the form
\begin{equation}
\label{eqn:HRG3d}
H(\u{A},\u{\varphi})=H_0(\u{A})+V(\u{\Omega}\cdot\u{A},\u{\varphi}),
\end{equation}
where $H_0$ is given by Eq.~(\ref{eqn:h0}),
combines four steps:\\
\indent \textbf{(1)} We shift the resonances ${\u \nu}_{k+1}\mapsto
{\u \nu}_{k}$: We require that the new angles $\u{\varphi}'$ satisfy
$$
\cos(\u{\nu}_{k+1}\cdot\u{\varphi})=
\cos(\u{\nu}_k\cdot\u{\varphi}'),$$
for $k\geq 1$.  This is performed
by the linear canonical transformation
$$
(\u{A},\u{\varphi})\mapsto (\u{A}',\u{\varphi}')=
(N^{-1}\u{A},\tilde{N}\u{\varphi}).
$$
Since $N$ is an integer matrix with determinant one, this
transformation preserves the ${\Bbb T}^3$-structure of the angles.  We
notice that the resonance $\u{\nu}_1$ is changed into
$\u{\nu}_0=(0,1,1)$ which satisfies the condition (\ref{eqn:cond}),
i.e.\ it is a non-resonant mode~: Some of the resonant modes are
turned
into non-resonant ones by this linear transformation.\\
This step changes the frequency ${\u \omega}_0$ into $\tilde{N}{\u
  \omega}_0= \sigma^{-1}{\u \omega}_0$ (since ${\u \omega}_0$ is an
eigenvector of $\tilde{N}$ by construction), and the vector
$\u{\Omega}$ into $\tilde{N}\u{\Omega}$. In order to keep a unit norm,
we define the image of $\u{\Omega}$ by
\begin{equation}
\label{eqn:map}
\u{\Omega}'=\frac{\tilde{N}\u{\Omega}}{\Vert \tilde{N}\u{\Omega} \Vert}.
\end{equation}
\indent \textbf{(2)} We rescale the energy (or equivalently time) by a
factor $\sigma$ (i.e.\ we multiply the Hamiltonian by $\sigma$), in
order to keep
the frequency fixed at ${\u \omega}_0$.\\
\indent \textbf{(3)} We rescale the actions~:
$$
H'({\u A},{\u \varphi})= \lambda H\left(\frac{{\u A}}{\lambda},{\u
    \varphi}\right),
$$
such that the mean-value of the coefficient of the quadratic term
in $H'$ is equal to $(\u{\Omega}'\cdot\u{A})^2/2$.  This normalization
condition is essential for the convergence of the transformation.
After Steps 1, 2 and 3, the Hamiltonian expressed in the new variables
is
\begin{equation}
H'(\u{A},\u{\varphi})=\lambda\sigma H\left(\frac{1}{\lambda}N\u{A},
\tilde{N}^{-1}\u{\varphi}\right).
\end{equation}
For $H$ given by Eq.\ (\ref{eqn:HRG3d}), this expression becomes
\begin{equation}
H'(\u{A},\u{\varphi})=\u{\omega}_0\cdot\u{A}+\frac{\sigma}{2\lambda}
\Vert \tilde{N}\u{\Omega}\Vert^2 (\u{\Omega}'\cdot\u{A})^2
+\lambda\sigma V\left(\frac{\Vert \tilde{N}\u{\Omega}\Vert}{\lambda}
\u{\Omega}'\cdot\u{A},\tilde{N}^{-1}\u{\varphi}\right).
\end{equation}
Thus the choice of the rescaling in the actions (Step 3) is
\begin{equation}
\lambda=\sigma \Vert \tilde{N}\u{\Omega}\Vert^2 (1+2\langle V^{(2)}\rangle),
\end{equation}
where $\langle V^{(2)}\rangle$ denotes the coefficient of the
mean-value of the quadratic part, in the representation~:
$$
V(\u{\Omega}'\cdot \u{A},\u{\varphi})=\sum_{j\geq 0} V^{(j)}(\u{\varphi})
(\u{\Omega}'\cdot\u{A})^j.
$$
\indent \textbf{(4)} We perform a near-identity canonical
transformation ${\mathcal U}_H$ that eliminates completely the
non-resonant part of the perturbation in $H'$. This transformation
satisfies the following equation~:
\begin{equation}
  \label{eq:uh}
  {\Bbb I}^- \left( H'\circ{\mathcal U}_H\right) =0,
\end{equation}
where ${\Bbb I}^-$ denotes the projection operator on the non-resonant
modes acting on a Hamiltonian $H$ as
\begin{equation}
{\Bbb I}^- H(\u{A},\u{\varphi})=\sum_{\u{\nu}\in I^-} H_{\u{\nu}}(\u{A})e^{i\u{\nu}\cdot
\u{\varphi}}.
\end{equation}
Equation~(\ref{eq:uh}) is solved by a Newton method.  We iterate a
change of coordinates as in KAM theory that reduces the non-resonant
modes of the perturbation from $\varepsilon$ to $\varepsilon^2$.  One
step of the elimination is performed by a Lie transformation
${\mathcal{U}}_S : ({\u A},{\u \phi})\mapsto ({\u A}',{\u \phi}')$,
generated by a function $S(\u{A},\u{\varphi})$.  The expression of a
Hamiltonian $H$ in the new coordinates is given by
\begin{equation}
\label{eqn:exp}
H\circ{\mathcal U}_S=\exp(\hat{S})H 
\equiv H+\{S,H\}+\frac{1}{2!}\{S,\{S,H\}\}+\cdots,
\end{equation}
where $\{ \, , \, \}$ is the Poisson bracket between two scalar
functions of the actions and angles:
\begin{equation}
\label{eqn:poiss}
\{f,g\}=\frac{\partial f}{\partial \u{\phi}} 
       \cdot \frac{\partial g}{\partial \u{A}} -
       \frac{\partial f}{\partial \u{A}} \cdot
       \frac{\partial g}{\partial \u{\phi}},
\end{equation}
and the operator $\hat{S}$ is defined as $\hat{S}H\equiv \{S,H\}$.
The generating function $S$ is chosen such that the order
$\varepsilon$ of the non-resonant part of the perturbation vanishes.
We construct recursively a series of Hamiltonians $H_n$, starting with
$H_1=H'$, such that the limit $H_{\infty}$ is canonically conjugate
to $H'$ but does not contain non-resonant modes, i.e.\ ${\Bbb I}^-
H_{\infty}=0$.  One step of this elimination procedure, $H_n\mapsto
H_{n+1}$, is done by applying a change of coordinates ${\mathcal U}_n$
such that the order of the non-resonant modes of
$H_{n+1}=H_n\circ{\mathcal U}_n$ is $\varepsilon_n^2$, where
$\varepsilon_n$ denotes the order of the non-resonant modes of $H_n$.
At the $n$-th step, the order of the non-resonant modes of $H_n$ is
$\varepsilon_0^{2^{n-1}}$, where $\varepsilon_0$ is the order of the
non-resonant modes of $H'$. If this procedure converges, it defines a
canonical transformation ${\mathcal U}_H={\mathcal U}_1\circ{\mathcal
  U}_2\circ\cdots\circ{\mathcal U}_n \circ\cdots$, such that the final
Hamiltonian $H_{\infty}=H'\circ{\mathcal U}_H$
does not contain any non-resonant mode.\\
The specific implementation of one step of this elimination procedure
will be described more explicitly in the next sections. We will
discuss two versions of this transformation~: The first one is a
renormalization for Hamiltonians in power series in the actions, and
the second one is a slightly different version which eliminates only
the non-resonant modes of the constant and linear part in the actions
of the rescaled Hamiltonian $H'$, and which allows us to define a
renormalization within a space of quadratic Hamiltonians in the
actions, following Thirring's approach of the KAM theorem~\cite{Bthir92}.\\
It has been proven in Ref.~\cite{koch99} that, for a sufficiently
small non-resonant part of the perturbation, the transformation
${\mathcal U}_H$ is a well-defined canonical transformation such that
a Hamiltonian expressed in these new coordinates does not have
non-resonant modes. The domain of definition of ${\mathcal U}_H$ has
been extended to some non-perturbative domain
in Ref.~\cite{koch99b}.\\
Concerning the quadratic case, we lack at this moment a theoretical
background to prove an analogous theorem. The convergence of the
elimination procedure in both cases outside the perturbative regime
($\varepsilon$ small) is
observed numerically.\\

In summary, the renormalization-group transformations we define act as
follows~: First, some of the resonant modes of the perturbation are
turned into non-resonant modes by a frequency shift and a rescaling.
Then, a KAM-type iteration eliminates these non-resonant modes, while
slightly changing the resonant modes.

\subsection{Renormalization scheme for Hamiltonians in power series 
  in the actions}
\label{sec:general}

We define in this section, one step of the elimination of the
non-resonant modes of the rescaled Hamiltonian $H'$. The
renormalization transformation acts on the following family of
Hamiltonians
\begin{equation}
  H(\u{A},\u{\varphi})=\u{\omega}_0\cdot\u{A}+\sum_{j=0}^{\infty} V^{(j)}(\u{\varphi})
  (\u{\Omega}\cdot\u{A})^j,
\end{equation}
with $\langle V^{(0)}\rangle =0$. We suppose that $\langle
V^{(2)}\rangle$ is nonzero (in order to have a twist direction in the
actions).  The approximations involved in this transformation are of
two types~: we truncate the Fourier series of the functions $V^{(j)}$,
i.e.\ we approximate a scalar function $f$ of the angles by~
   \begin{equation}
   \label{eqn:coeff}
   f^{[\leq L]}(\u{\phi})=\sum_{\u{\nu}\in{\mathcal C}_L} f_{\u{\nu}} 
   e^{i\u{\nu}\cdot\u{\phi}},
   \end{equation}
   where ${\mathcal C}_L=\{\u{\nu}=(\nu_1,\nu_2,\nu_3)\in {\Bbb
     Z}^3\left| \max_i |\nu_i|\leq L\right.\}$, and we also neglect
   all the terms of order $O\left((\u{\Omega}\cdot
     \u{A})^{J+1}\right)$ that are produced by
   the transformation.\\
   One step of the elimination procedure $H\mapsto \tilde{H}=H\circ
   {\mathcal U}_S$ is constructed as follows~: We consider that
   $V^{(j)}$ depends on a small parameter $\varepsilon$, such that
   ${\Bbb I}^-V^{(j)}$ is of order $\varepsilon$. We define $H_0$, the
   integrable part of $H$ as
   \begin{equation}
   H_0(\u{A})=\u{\omega}_0\cdot\u{A}+\langle V^{(2)}\rangle
   (\u{\Omega}\cdot\u{A})^2,
   \end{equation}
   and the perturbation of $H_0$ is denoted $V'=H-H_0=V-\langle
   V^{(2)}\rangle (\u{\Omega}\cdot\u{A})^2$.  In order to eliminate
   the non-resonant modes of $V^{(j)}$ to the first order in
   $\varepsilon$, we perform a Lie transformation $ {\mathcal U}_S:
   (\u{A},\u{\varphi})\mapsto (\u{A}',\u{\varphi}')$ generated by a
   function $S$ of order $\varepsilon$ and of the form
   \begin{equation}
   S(\u{A},\u{\varphi})=i\sum_{j=0}^J Y^{(j)}(\u{\varphi})(\u{\Omega}
   \cdot\u{A})^j +a\u{\Omega}\cdot\u{\varphi}.
   \end{equation}
   The first terms of $\tilde{H}$ are
   $H_0+V'+\{S,H_0\}+\{S,V'\}+O(\varepsilon^2)$.  The function $S$ is
   determined by imposing that the order $\varepsilon$ vanishes~:
   \begin{equation}
     \label{eq:detS}
   {\Bbb I}^-\{S,H_0\}+{\Bbb I}^-V'+{\Bbb I}^-\{S,{\Bbb I}^+V'\}=0.
   \end{equation}
   This condition determines the non-resonant modes of $S$. For the
   resonant ones, we choose ${\Bbb I}^+S=0$. Thus we notice that the
   mean value of $\{S,{\Bbb I}^+V'\}$ is zero.  The constant $a$
   eliminates the linear term in the
   $(\u{\Omega}\cdot\u{A})$-variable, $\langle V^{(1)} \rangle$, by
   requiring that $\langle \{S,H_0\}\rangle +\langle V^{(1)}
   \rangle\u{\Omega}\cdot\u{A}=0$:
   \begin{equation}
   a=-\frac{\langle V^{(1)}\rangle}{2\Omega^2\langle V^{(2)}\rangle}.
   \end{equation}
   We solve Eq.~(\ref{eq:detS}) by a Newton method with an initial
   condition which satisfies ${\Bbb I}^-\{S,H_0\}+{\Bbb I}^- V=0$,
   since ${\Bbb I}^+V'$ is expected in general to be small. Then we
   compute $\tilde{H}= H\circ{\mathcal U}_S$ by calculating
   recursively the Poisson brackets
   $\hat{S}^kH=\hat{S}\hat{S}^{k-1}H$, for $k\geq 1$.  Denoting
   $H_k=\hat{S}^kH$, $\tilde{H}$ becomes
   \begin{equation}
   \tilde{H}=\sum_{k=0}^{\infty}\frac{H_k}{k!}.
   \end{equation}
 
   \subsection{Thirring's scheme for quadratic Hamiltonians}\label{sec:quadra}
   
   We consider the following family of quadratic Hamiltonians in the
   actions and described by three scalar functions of the angles
  \begin{equation}
   H({\u A},{\u \varphi})=\u{\omega}_0\cdot \u{A}+
    m({\u \varphi})( {\u \Omega}\cdot{\u A})^{2} 
       + g({\u \varphi}){\u \Omega}\cdot{\u A}
       + f({\u \varphi}) \label{eqn:Hquadra}  ,
  \end{equation}
  The KAM transformations are constructed such that the iteration
  stays within the space of Hamiltonians quadratic in the
  actions~\cite{Bthir92}.  In order to prove the existence of a torus
  with frequency vector ${\u \omega}_0$ for Hamiltonian systems
  described by Eq.\ (\ref{eqn:Hquadra}), it is not necessary to
  eliminate $m$, but only $g$ and $f$ (the main point is that the
  torus with frequency vector $\u{\omega}_0$ is located at
  $\u{A}=\u{0}$ for any $H$ with $f=g=0$, even if $H$ is not globally
  integrable).  The elimination of $f$ and $g$ can be achieved with
  canonical transformations with generating functions that are linear
  in the action variables, and thus map the family of Hamiltonians
  (\ref{eqn:Hquadra}) into itself.  This is very convenient
  numerically, as one only works with three scalar functions $m$, $g$
  and $f$. The only approximation involved in the numerical
  implementation of the transformation is a truncation of
  the Fourier series of these functions, according to Eq.~(\ref{eqn:coeff}).\\
  In this section, we describe one step of the elimination of the
  non-resonant modes $H\mapsto \tilde{H}=H\circ{\mathcal U}_S$.  We
  assume that $g$ and $f$ depend on a (small) parameter $\varepsilon$,
  in such a way that ${\Bbb I}^-g$ and ${\Bbb I}^-f$ are of order
  $\varepsilon$. The idea is to eliminate the non-resonant modes of
  $g$ and $f$ to first order in $\varepsilon$, at the expense of
  adding terms that are of order $O(\varepsilon)$ in the resonant
  modes and of order $O(\varepsilon^2)$ in the non-resonant modes.
  This is performed by a Lie transformation $ {\mathcal U}_S : ({\u
    A},\u{\varphi})\mapsto(\u{A}', \u{\varphi}')$ generated by a
  function $S$ of order $\varepsilon$ linear in the action variables,
  of the form
  \begin{equation}
  \label{eqn:S}
  S(\u{A},\u{\varphi})=Y(\u{\varphi})\u{\Omega}\cdot\u{A}
  +Z(\u{\varphi})+a\u{\Omega}\cdot\u{\varphi}\ ,
  \end{equation}
  characterized by two scalar functions $Y$, $Z$, and a constant $a$.
  The expression of the Hamiltonian in the new variables is obtained
  by Eq.~(\ref{eqn:exp}).  A consequence of the linearity of $S$ in
  $\u{A}$ is that the Hamiltonian $\tilde{H}$ is again quadratic in
  the actions, and of the form
  \begin{equation}
  \tilde{H}({\u A},{\u \varphi})={\u \omega}_0\cdot\u{A}+
                \tilde{m}({\u \varphi})( {\u \Omega}\cdot{\u A})^{2}
                + \tilde{g}({\u \varphi}){\u \Omega} \cdot{\u A}
                + \tilde{f}({\u \varphi}).\label{image}
  \end{equation}
  This can be seen by this simple argument~: Given a quadratic
  function in the $(\u{\Omega}\cdot\u{A})$-variable and a linear
  function $S$, then $\hat{S}H$ is again quadratic in the
  $(\u{\Omega}\cdot\u{A})$-variable. The derivatives $\partial H
  /\partial \u{A}$ and $\partial S/\partial \u{\varphi}$ are linear in
  the $(\u{\Omega}\cdot\u{A})$-variable, and $\partial H/\partial
  \u{\varphi}$ is quadratic, while $\partial S/\partial \u{A}$ is
  constant in this variable.  Therefore, $\hat{S}H$ given by
  Eq.~(\ref{eqn:poiss}) is quadratic.  Consequently, iterating this
  argument, $\exp(\hat{S})H$ is also quadratic.  We notice that the
  vector $\u{\Omega}$ remains unchanged during each step of the
  elimination.\\ The functions $Y$, $Z$, and the constant $a$ are
  chosen in such a way that ${\Bbb I}^-\tilde{g}$ and ${\Bbb
    I}^-\tilde{f}$ vanish to order $\varepsilon$.  The constant $a$
  corresponds to a translation in the actions, which has the purpose
  of eliminating the mean value of the linear term in the
  $(\u{\Omega}\cdot\u{A})$-variable.  Then we express $\tilde{H}$ by
  calculating recursively the Poisson brackets
  $\hat{S}^kH=\hat{S}\hat{S}^{k-1}H$,
  for $k \geq 1$, like in the previous section.\\
  We notice that for some purposes it is more convenient to eliminate
  also the non-resonant part of $m$, together with the one of $g$ and
  $f$ (see the remarks in Refs~\cite{koch99,chan98b}). But such
  elimination procedure generates arbitrary orders in the
  $\u{\Omega}\cdot\u{A}$-variable and this leads to the first version
  of the transformation.  The advantage to work with this
  second version of the elimination procedure is that the
  Hamiltonians~(\ref{eqn:Hquadra}) are described by only three scalar
  functions of the angles. Thus it is numerically more efficient since
  the renormalization map is of lower dimension. Concerning the
  renormalization transformation for Hamiltonians in power series in
  the actions, we truncate the Fourier series of each scalar function
  of the angles with a cut-off parameter $L$, and the Taylor series
  with a cut-off parameter $J$. For fixed $L$ and $J$, the dimension
  of the renormalization map is equal to $(J+1)(2L+1)^3+2$. Concerning
  the renormalization defined for quadratic Hamiltonians, the
  renormalization map is of dimension $3(2L+1)^3+2$.\\
  {\em Remark :} Another advantage to work the Thirring's version of
  the renormalization is that it can be generalized more easily to
  non-degenerate Hamiltonians
  $H(\u{A},\u{\phi})=H_0(\u{A})+V(\u{A},\u{\phi})$ with $\mbox{Rank }
  \partial^2 H_0 /\partial \u{A}^2 \geq 1$. (see Ref.~\cite{chan99b}).


\section{Determination of the critical coupling}
\label{sect:result}

We consider the following quasi-periodically driven pendulum model~:
\begin{equation}
  \label{eq:fp}
  H=\frac{1}{2}p^2-p +\varepsilon \left( \cos x +\cos
  (x+\nu_1t)+\mu \cos(x+\nu_2 t)\right),
\end{equation}
with the frequencies $\nu_1=\sigma+1$, $\nu_2=\sigma^2+1$, and $\mu$
a real parameter.
This model can be mapped into the following degenerate Hamiltonian
system with three degrees of freedom~:
\begin{equation}
\label{eqn:param}
H_\varepsilon({\u A},{\u \varphi})={\u \omega}_{0}\cdot{\u A}+
\frac{1}{2}({\u \Omega}\cdot{\u A})^{2}
+\varepsilon f({\u \varphi}) \ ,
\end{equation}
where $\u{\Omega}=(1,1,-1)$ and the perturbation $f$ is given by
\begin{equation}
\label{eq:pertu}
f({\u \varphi})=\mu \cos\varphi_1+ \cos\varphi_2+ \cos\varphi_3, 
\end{equation}
This can be seen by considering the three angles $\varphi'_1=x$,
$\varphi'_2=\nu_1 t \mbox{ mod } 2\pi $, and $\varphi'_3=\nu_2 t \mbox{
  mod } 2\pi$. The Hamiltonian~(\ref{eq:fp}) becomes~:
\begin{equation}
  \label{eq:hamint}
  H=\frac{1}{2}A_1^{'2}-A'_1+\nu_1 A'_2+\nu_2 A'_3+\varepsilon \left( \cos \varphi'_1 
+\cos(\varphi'_1+\varphi'_2)+\mu \cos(\varphi'_1+\varphi'_3)\right),
\end{equation}
where we added $\nu_1 A'_2+\nu_2 A'_3$ to the Hamiltonian in order to satisfy the
equations of motion for the new variables $\varphi'_2$ and
$\varphi'_3$ when $\varphi'_2=\varphi'_3=0$ at time $0$.
The linear canonical transformation
$(\u{A}',\u{\phi}')=(M\u{A},\tilde{M}^{-1}\u{\phi})$ with
$$
M=\left( \begin{array}{ccc} 1 & 1 & -1 \\ 0 & 1 & 0\\ 1 & 0 &
    0\end{array} \right),
$$
maps Hamiltonian~(\ref{eq:hamint}) into Hamiltonian~(\ref{eqn:param}).

\subsection{Torus with frequency vector $\u{\omega}_0$}

We are looking at the break-up of the invariant torus with frequency
vector $\u{\omega}_0$ which is located at $\u{\Omega}\cdot\u{A}=0$ for
Hamiltonian~(\ref{eqn:param}) with $\varepsilon=0$.  We notice that
this torus is located at $p=0$ for Hamiltonian~(\ref{eq:fp}) and its
frequency is equal to $-1$. The three main resonances (of order
$\varepsilon$) are located at $p=1$, $p=-\sigma$, and $p=-\sigma^2$,
and thus the torus is located in between the resonances $p=-\sigma$
and $p=1$. We first discuss two rough estimates of the critical
coupling obtained by some drastic simplifications~: $(a)$
Applying Chirikov's criterion~\cite{chir79} gives the
following approximate value for the critical coupling 
$\varepsilon_c \approx
(\sigma+1)^2/16 \approx 0.3377$. $(b)$ If we neglect the effect
of the resonance located at $p=-\sigma^2$, e.g., by setting $\mu=0$ in the
perturbation of Eq.~(\ref{eq:fp}), we can apply the renormalization
procedure described in Ref.~\cite{chan00a} for Hamiltonians with two
degrees of freedom. Since Hamiltonian~(\ref{eq:hamint}) does not depend
on $\varphi'_3$ in this case, $A'_3(t)$ is constant and the problem
reduces to the study of the break-up of the invariant torus with
frequency vector $\u{\omega}_0^{(2d)}=(-1,\sigma+1)$ for the following
Hamiltonian system with two degrees of freedom~:
$$
H=\frac{1}{2} A_1^{'2} +\u{\omega}_0^{(2d)}\cdot (A'_1,A'_2) +\varepsilon
(\cos\varphi'_1+\cos (\varphi'_1+\varphi'_2)).
$$
This method gives $\varepsilon_c\approx
0.112$.\\

\begin{figure}[ht] \vspace{0.4cm} \hspace{0.cm}
\unitlength 1cm
\begin{picture}(10,10)(0,10)
\put(0,18){\centerline{ \includegraphics*[scale=0.4,angle=-90]{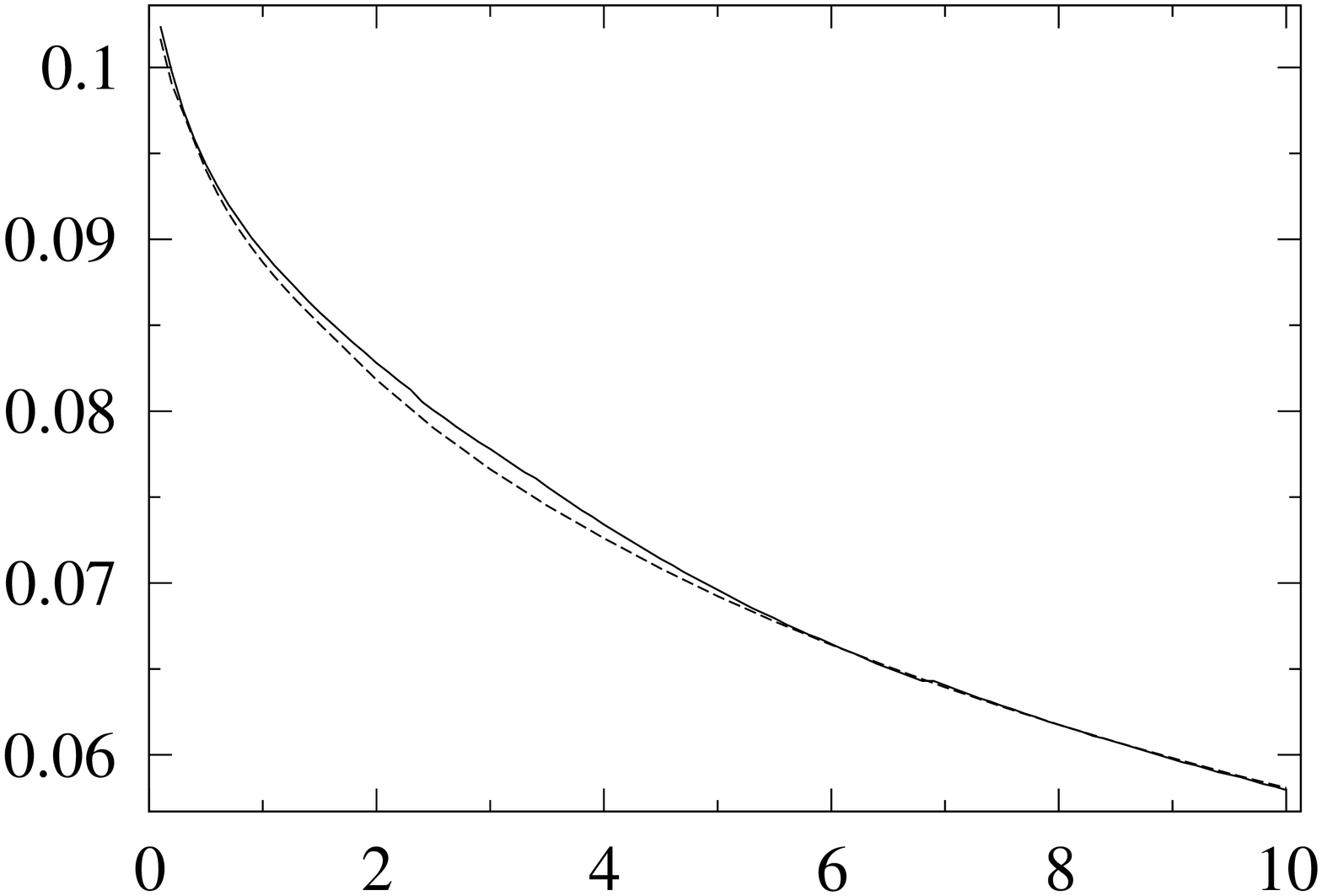}}}
\Large
\put(6.8,9.8){$\mu$}
\put(0,13.5){$\varepsilon_c(\mu)$}
\end{picture}
 \caption{Values of the critical couplings $\varepsilon_c(\mu)$ for
Hamiltonian~(\ref{eq:fp}) with $\mu\in [0,10]$. The dashed curve
   is obtained by the renormalization method, and the continuous curve is 
   obtained by Frequency Map Analysis.} \label{fig:mu}
\end{figure}

We perform the complete renormalization procedure described in the
preceding sections for Hamiltonian system~(\ref{eqn:param}) for a
given value of $\mu$. We fix the cut-off parameters
$L$ and $J$ of the renormalization transformation, and we take
successively larger couplings $\varepsilon$ in order to determine
whether the renormalization transformation ${\mathcal R}$ converges to
an integrable Hamiltonian, or whether it diverges.  By a bisection
procedure, we determine the critical coupling $\varepsilon_c(\mu ; L,J)$
such that as $n$ tends to $+\infty$~:
\begin{eqnarray*}
  && {\mathcal R}^nH_{\varepsilon}\to H_0 \quad \mbox{ for }
  |\varepsilon|< \varepsilon_c,\\
  && {\mathcal R}^nH_{\varepsilon}\to \infty \quad \mbox{ for }
  |\varepsilon|> \varepsilon_c.
\end{eqnarray*}
Table 1 gives the values of $\varepsilon_c$ for $\mu=1$ as a function of $L$ and
$J$ computed by the two renormalization transformations. We notice
that the values $\varepsilon_c(\mu=1 ; L,J)$ converge to $\varepsilon_c\approx
0.0886$ as $L$ and $J$ grow.\\
Figure~\ref{fig:mu} shows the values of $\varepsilon_c(\mu)$ for $\mu
\in [0,10]$ determined by the renormalization transformation for
quadratic Hamiltonians with cut-off parameter $L=3$.

\subsection{Frequency Map Analysis}

\begin{figure}[ht] \vspace{0.4cm} \hspace{0.cm}
  \centerline{ \includegraphics*[scale=0.6]{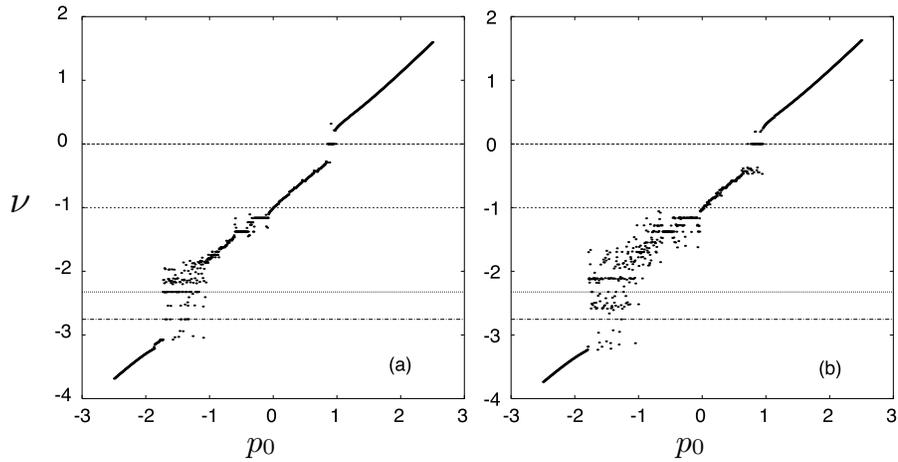}}
 \caption{Frequency map for Hamiltonian~(\ref{eq:fp}) with $\mu=1$ and
   with $\epsilon = 0.09$ (a) and $\epsilon = 0.12$ (b). The
   dotted lines correspond to $\nu = 0, -1,-1 -\sigma,-1 -\sigma^2$.
   $T = 1000$.} \label{fig:fmafig1}
\end{figure}

Frequency Map Analysis\cite{lask90,lask92,lask93,lask99} allows to
study the destruction of KAM tori by looking at the regularity of the
frequency map defined from the action like variables to the
numerically determined frequencies. In the present case (\ref{eq:fp}),
the frequency map is very simple as the system is equivalent to a one
degree of freedom system with a quasiperiodic perturbation
\cite{lask99}.  The angle variable $x$ can then be fixed to $x=0$, and
we are left to a one dimensional map $F:R \longrightarrow R$, $p_0
\longrightarrow \nu $, where $\nu$ is determined numerically from the
output of the numerical integration of $(p(t), x(t))$ over a time
interval of length $T$, starting with initial conditions $(p(0),x(0))
=(p_0,0)$ \cite{lask99}.  On the set of KAM tori $F$ is regular, or
more precisely, it can be extended to a smooth function. Thus, when $F$
appears to be non regular, this is an indication that all tori are destroyed
in the corresponding interval. This allows to obtain a global
vision of the dynamics of the system, as illustrated by
figure~\ref{fig:fmafig1} for Hamiltonian~(\ref{eq:fp}) with $\mu=1$,
where $\nu = F(p_0)$ is plotted versus $p_0$ for $\epsilon = 0.09$ and
$\epsilon = 0.12$, and for $p_0 \in [-2.5,2.5]$ for a moderate
precision ($T=1000$).  It seems clear on these figures that in the
region between the two resonances $\nu = -1-\sigma^2$ and $\nu=0$,
there are no tori left for $\epsilon =0.12$, while many of them
remain in the region $\nu \in [-1,0]$ for $\epsilon = 0.09$.

\begin{figure}[ht] \vspace{0.4cm} \hspace{0.cm}
  \centerline{ \includegraphics*[scale=0.6]{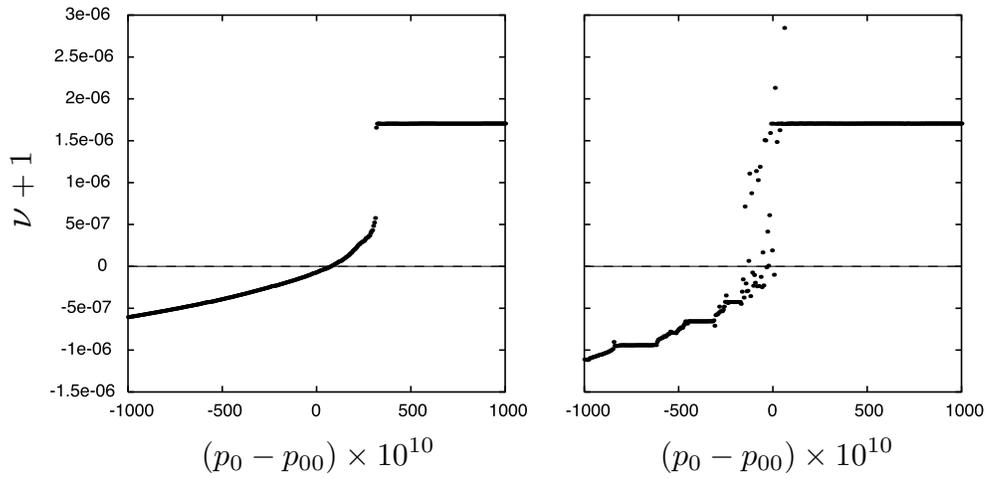} }
\caption{Frequency map for Hamiltonian~(\ref{eq:fp}) with $\mu=1$ and
  with parameter $\epsilon = 0.089$ (a) and $\epsilon = 0.0895$ (b).
  $\nu+1$ is plotted versus $(p_0 - p_{00})\times 10^{10}$, where
  $p_{00}= -3.3307 \times 10^{-4}$. The dotted lines corresponds to
  $\nu = -1$. $T = 500 000$.} \label{fig:fmafig2}
\end{figure}

In order to have a more detailed view for the destruction of the tori
with frequency $\nu=-1$, we have extended the time interval to $T=500
000$, and reduced very much the stepsize in $p_0$ (figure~\ref{fig:fmafig2}). With
these settings and for $\mu=1$, it appears clearly that the torus with frequency
$\nu=-1$ is destroyed for $\epsilon = 0.0895$ (b), while the behavior
of the frequency map appears to be very regular for $\epsilon = 0.089$
(a). It should be noted that the frequency map analysis provides a
criterion for the destruction of tori. The fact that the frequency curve
appears to be irregular provides an evidence that the tori are destroyed,
but when the curve is regular, a higher accuracy (which means a longer
time span $T$) could reveal the destruction of additional tori.

For $\epsilon = 0.0895$ and for $\mu=1$, all tori in the vicinity of $\nu=-1$ are
destroyed, which is in agreement with the value $\epsilon_c \approx
0.0886$ found with the renormalization technique, as this is the
largest value of $\epsilon$ for which the renormalization converges.\\
We have performed computations for other values of the parameter
$\mu$. Figure~\ref{fig:mu} shows the agreement between the couplings
obtained by Frequency Map Analysis and the ones obtained by the
renormalization transformation. A better accuracy can be obtained by taking a 
larger cut-off parameter $L$ for the renormalization computations.

\subsection{Last KAM torus} 

It appears clearly in figure~\ref{fig:fmafig1} that the torus with frequency $\nu=-1$
is not the last torus to be destroyed in the interval of frequencies
$[-1-\sigma^2,0]$ for $\mu=1$, and that many tori still survive for $\epsilon >
0.0895$.  We thus have searched for the value of the parameter
$\epsilon$ for which the last torus disappears.  This is done by
increasing the value of the parameter $\epsilon$ until all tori
disappear.  In this procedure, at each stage we identify the
resonant islands and chaotic regions, and then decrease the stepsize
in $p_0$.  This allows one to obtain very refined details as one gets
close to the critical value of $\epsilon$.

\begin{figure}[ht] \vspace{0.cm} \hspace{0.cm}
  \centerline{ \includegraphics*[scale=0.8]{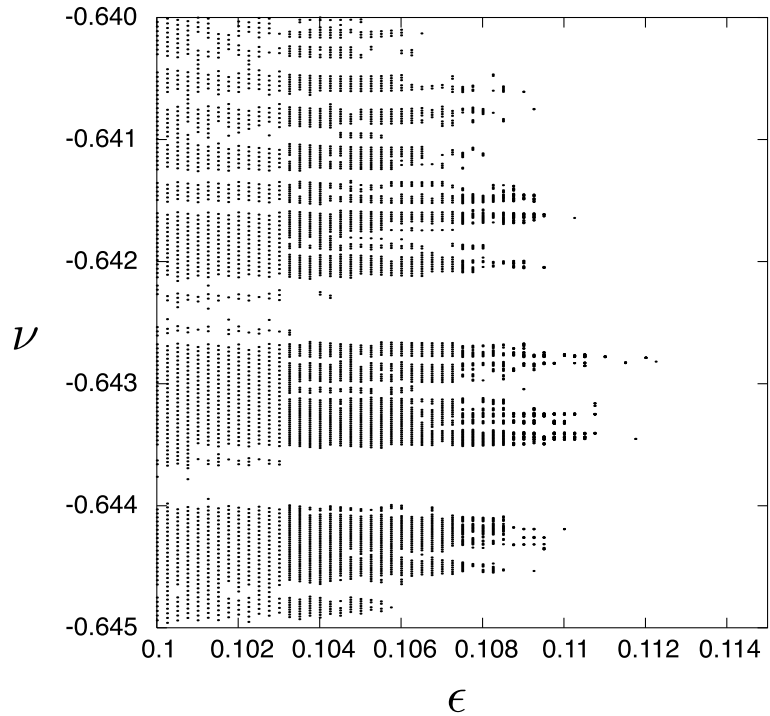} }
\caption{Determination of the last KAM tori for the Hamiltonian
  (\ref{eq:fp}) with $\mu=1$.
  For each selected value of the parameter $\epsilon$, the set of
  frequencies $\nu$ for which tori are not destroyed is plotted versus
  $\epsilon$. The critical value of $\epsilon$ for which no tori will
  survive is thus $\epsilon_c\approx 0.11$.} \label{fig:fmafig3}
\end{figure}

The frequencies of the last invariant torus can be investigated by
frequency map analysis. This method gives a critical threshold at
about $\varepsilon_c \approx 0.11$ (see figure~\ref{fig:fmafig3}). For
Hamiltonian~(\ref{eq:fp}) with $\mu=1$, the last invariant torus is located 
between the resonances $p=1$ and $p=-\sigma$ at $p_0 \approx 0.354$.
The corresponding frequencies for the last three-dimensional torus for
the family of Hamiltonians~(\ref{eqn:param}) are $p_0+\sigma^2$,
$p_0+\sigma$ and $1-p_0$.

A nearby invariant torus has the frequencies $(\sigma +2)(1-p_0)$,
$(\sigma^{-2}+2)(1-p_0)$ and $1-p_0$. The critical coupling for the
break-up of this invariant torus can be computed by the
renormalization transformation defined in Sec.~\ref{sec.2}, mainly by
defining first a unimodular transformation that maps the frequencies
of the torus into the frequencies $\sigma^2$, $\sigma$ and 1 (the
perturbation is then different from Eq.~(\ref{eq:pertu})). The
renormalization gives $\varepsilon_c \approx 0.11$.\\
If we neglect the resonance located at $p=-\sigma^2$, renormalization
methods and frequency map analysis show that there are two last KAM
tori, one located at $p\approx 0.112$ and the other one at $p\approx
-0.436$, with a critical threshold of $\varepsilon_c \approx 0.149$.
Thus the effect of the third resonance at $p=-\sigma^2$ is to
destabilize the motion in the region between $p=1$ and $p=-\sigma$
closer to $p=-\sigma$ (since the third resonance is closer to
$p=-\sigma$).  Consequently the last KAM torus is located nearer $p=1$
than in the system without the third resonance, and the value of the
threshold of global stochasticity (break-up of the last KAM surface)
is smaller.

\section*{Acknowledgments}
We acknowledge useful discussions with G.\ Gallavotti, H.\ Koch, and
R.S.\ MacKay.  Support from EC Contract No.\ ERBCHRXCT94-0460 for the
project ``Stability and Universality in Classical Mechanics'' is
acknowledged. CC thanks support from the Carnot Foundation.


\newpage

\begin{table}[ht]
\begin{center}
\begin{tabular}{|c| c c c c c c|}
\hline
  & \multicolumn{5}{c}{RG1} & RG2\\
\hline
  L &  J=2     &   J=3    &   J=4    &   J=5    &   J=6    &          \\
  \hline
  2 & 0.089230 & 0.090104 & 0.089924 & 0.089936 & 0.08988  & 0.089114 \\
  3 & 0.087744 & 0.088466 & 0.088438 & 0.088379 & 0.088326 & 0.088673 \\
  4 & 0.087672 & 0.088392 & 0.088283 & 0.088194 & 0.088238 & 0.088645 \\
  5 & 0.087667 & 0.088384 & 0.088234 & 0.088184 & 0.088224 & 0.088649 \\
  6 & 0.087666 & 0.088384 & 0.088237 & 0.088186 & 0.088226 & 0.088646
  \\
  15 & - & - & - & - & - & 0.088644\\
\hline
\end{tabular}
\end{center}
\caption{Critical coupling $\varepsilon_c$ for
  Hamiltonian~(\ref{eq:fp}) with $\mu=1$ as a function of the
  cut-off parameters $L$ and $J$, computed with the transformation RG1 defined in 
Sec.~\ref{sec:general}, and with the transformation RG2 defined in
Sec.~\ref{sec:quadra}. Frequency Map Analysis indicates that the
spiral mean torus is broken for couplings larger than 0.0895.}
\end{table}

\newpage

\newpage

\end{document}